\documentclass[a4paper,11pt]{article}
\usepackage{amssymb}

\newcommand{\qeda}{\hspace{10mm}\hfill $\square$}

\newtheorem{theorem}{Theorem}
\newtheorem{corollary}{Corollary}
\newtheorem{lemma}{Lemma}
\newtheorem{proposition}{Proposition}

\title{Blow-up rate for a semi-linear accretive wave equation}
\author{H. Faour\footnote{Cermics, Paris-Est-ENPC,
ParisTech, 6 et 8 avenue Blaise Pascal, Cit\'e Descartes
Champs-sur-Marne, 77455 Marne-la-Vall\'ee Cedex 2, France. E-mail:
faourh@cermics.enpc.fr.\newline \indent ${}^2$ \hskip-.1cm M.
Jazar, LaMA-Liban, AZM Research Center, Lebanese University, P.O.
Box 37, Tripoli Lebanon. E-mail: mjazar@ul.edu.lb.\newline \indent
${}^3$ \hskip-.1cm Ch. Messikh, Universit\'e Badji Mokhtar,
D\'epartement de Math\'ematiques BP 12 - 23000, Annaba, Alg\'erie.
Email: messikh\_chahrazed@yahoo.fr  \newline The second author is
supported by a grant from Lebanese National Council for Scientific
Research. This work is partially supported by the project ANR MICA
(2006-2010).}, \hskip .2cm M. Jazar${}^2$ \hskip .2cm and Ch.
Messikh${}^3$}
\date{}

\begin{document}

\maketitle

\noindent{\small{\textbf{Abstract.} In this paper we consider the
semi-linear wave equation: $u_{tt}-\Delta u=u_t|u_t|^{p-1}$ in
$\mathbb{R}^N$. We provide an associated energy. With this energy
we give the blow-up rate for blowing up solutions in the case of
bounded below energy.}}

\noindent{\small\textbf{AMS Subject Classifications:}}
{\small{35L05,35L67}}\hfill

\medskip

\noindent{\small\textbf{Keywords:}} {\small{ Wave equation, finite
time blow-up, blow-up rate.}}

\section{Introduction}

We consider the following semi-linear wave equation:
\begin{equation}\label{e1.1}
\left\{
\begin{array}{ll}
u_{tt}-\Delta u=u_t\left| u_t\right| ^{p-1}& t\in[0,T),\, x \in
\mathbb{R}^N\\ &\\ (u(x,0), u_t( x,0)) =(u_0, u_1)\in Y_{loc,
u}(\mathbb{R}^N),&
\end{array}
\right.
\end{equation}
where $Y_{loc, u}$ is either $Y_{loc, u}^{2,2}:=
H_{loc,\,u}^2(\mathbb{R}^N)\times H_{loc,\,u}^1(\mathbb{R}^N)$ or
$Y_{loc, u}^{2,\infty}:= (W_{loc,\,u}^{2,\infty}\cap
H_{loc,\,u}^2(\mathbb{R}^N))\times (L_{loc,\,u}^{\infty}\cap
H_{loc,\,u}^1(\mathbb{R}^N))$, with
$$
L_{loc,\,u}^2\left( \mathbb{R}^N\right) =\left\{
v\colon\mathbb{R}^N\rightarrow \mathbb{R};\;\| v\|
_{L_{loc,\,u}^2}:=\sup_{x_{0}\in \mathbb{R}^N}\int_{| x-x_{0}|
\leq 1}| v(x)|^2dx\;<\infty \right\}
$$
and
$$H_{loc,\,u}^1\left( \mathbb{R}^N\right):=\left\{ v\in
L_{loc,\,u}^2\left( \mathbb{R}^N\right); \;\left| \nabla v\right|
\in \;L_{loc,\,u}^2\left( \mathbb{R}^N\right) \right\},$$ and
corresponding definitions for $L_{loc,u}^{\infty}$ and
$W_{loc,u}^{k,\infty}$. We assume in addition that
\begin{equation}\label{hypp}
\left\{
\begin{array}{ll}(u_0,u_1)\in Y_{loc, u}^{2,2}\mbox{ and }
    \left|\begin{array}{ll}

        p\in \mathbb{N}\cap\left(1,\frac N{N-2}\right)&\mbox{if } N\ge 3\\
                   \mbox{or} & \\
        p \in \{2,3,\cdots\}& \mbox{ if } N\le2.
    \end{array}\right. \\
&\\
(u_0,u_1)\in Y_{loc, u}^{2,\infty}, \,N\le 3,\mbox{ and
}p\in\mathbb{R}\cap(1,+\infty)\end{array}\right.
\end{equation}

All of this is due to the local existence result (see for instance
Faour, Fino and Jazar \cite{FaourFinoJazar}).

A very rich literature has been done on the semi-linear equation
\begin{equation}\label{e1.1p}
u_{tt}-\Delta u=au_t\left| u_t\right| ^{p-1}+bu|u|^{q-1}
\end{equation}
with $a$, $b$, $p$ and $q$  are real numbers, $p,q\ge1$. When
$a\le 0$ and $b=0$ then the damping term $au_t|u_t|^{p-1}$ assume
global existence in time for arbitrary data (see, for instance,
Harraux and Zuazua \cite{HarauxZuazua} and Kopackova
\cite{Kopackova}). When $a\le 0$, $b>0$ and $p>q$ then one can
cite, for instance, Levine \cite{Levine1974} and Georgiev and
Todorova \cite{GeorgievTodorova}, that show the existence of
global solutions (in time) under negative energy condition. When
$a\le 0$, $b>0$ and $q>p$, or when $a\le 0$, $b>0$ and $p=1$ then
one can cite \cite{GeorgievTodorova} and Messaoudi
\cite{Messaoudi} where they show finite time blowing up solutions
under sufficiently large negative energy of the initial condition.

The first to consider the case $a>0$ was Haraux \cite{Haraux1992}
(with $b=0$ on bounded domain), who construct blowing up solutions
for arbitrary small initial data. See also Jazar and Kiwan
\cite{JazarKiwan1} and the references therein for the same
equation (\ref{e1.1}) on bounded domain. We refer to Levine, Park
and Serrin \cite{LevineParkSerrin} and the references therein for
the whole space-case $\mathbb{R}^N$. Finally, we refer to Haraux
\cite{Haraux1992}, Souplet \cite{Souplet1995, Souplet1998} and
Jazar and Souplet \cite{JS} concerning the ODE case.

Unlike previous work where the considered question was to provide
conditions ensuring finite time blowup for the solution, recent
interesting work has been done aiming at understanding the
behavior of blowing up solutions in $H^1_{loc,u}\times
L^2_{loc,u}$-norm. This was the aim of the paper of Antonini and
Merle \cite{AntoniniMerle} and also the series of papers of Merle
and Zaag \cite{MerleZaag2003, MerleZaag2005MA, MerleZaag2005IMRN}
where they was concerned by the blow-up rate for (\ref{e1.1p}) in
the case $a=0$ and $b=1$. They showed that the blow-up rate is
that of the associated ODE ($u''=u^p$) for $1<p\le 1+\frac4{N-1}$,
and in \cite{MerleZaag2005IMRN} they study the growth rate near
the blowup surface.

In this paper we consider the case $a=1$ and $b=0$.

For the rest of the paper, and following
\cite{AntoniniMerle,MerleZaag2003, MerleZaag2005MA,
MerleZaag2005IMRN}, we consider solutions $u$ of (\ref{e1.1}) that
blow-up in finite time $T>0$ in the space $Y_{loc,\,u}( \mathbb{R}
^N)$. Our aim is to study the blow-up behavior of $u(t)$ as
$t\uparrow T$. We compare the growth of $u_t$ and $k$, the
solution of the simplest associated ODE: $k_{tt}=k_t|k_t|^{p-1}$.
Nevertheless, the presence of the force term $u_t|u_t|^{p-1}$
makes the work more complicated. To remedy this difficulty, and
inspired by the work of Rivera and Fatori \cite{RiveraFatori}, we
rewrite (\ref{e1.1}) as
\begin{equation}\label{e1.2}
\left\{
\begin{array}{ll}
u_{tt}-\int_{0}^t\Delta u_t\left( \tau \right) d\tau -\Delta
u_{0}=u_t\left| u_t\right| ^{p-1},&t\in[0,T),\,x\in
\mathbb{R}^N,\\ &\\ (u(x,0), u_t(x,0)) =(u_0(x), u_1(x)) \in
Y_{loc,\,u}.&
\end{array}
\right.\end{equation} Then, putting
\begin{equation}\label{e1.3}
v\left( x,t\right) =u_t\left( x,t\right)
\end{equation}
in (\ref{e1.2}), we obtain the following integro differential PDE
\begin{equation}\label{e1.4}
\left\{
\begin{array}{ll}
v_t-\int_{0}^t\Delta v\left( \tau \right) d\tau -\Delta
u_{0}\left( x\right) =v\left| v\right| ^{p-1}&t\in[0,T),\,x\in
\mathbb{R}^N,\\ & \\ v\left( x,0\right) =u_1\left(
x\right)=:v_{0}.&
\end{array}
\right.
\end{equation}

Now, we introduce (see \cite{Alinhac,AntoniniMerle}) the following
change of variables. For $a\in \mathbb{R}^N$ and $T'>0$, with
$\beta:=\frac1{p-1}$,
\begin{equation}\label{e1.5}
\begin{array}{lll}
z=x-a,&s=-\log \left(T'-t\right),&v\left( t,x\right) =\left( T'
-t\right) ^{-\beta }\theta_{T' ,\;a}\left( s,z\right)
\end{array}
\end{equation}
and
$$\left( T' \right) ^{\beta+1}u_0(x)=:
\theta_{a,00}\hskip 1cm \left( T' \right)^{\beta}v(x,0) =: \theta(
s_{0},z) =:\theta_{a,0} (z)$$ where $s_0:=-\log T'$. We then see
that the function $\theta_{T,a}$ (we write $\theta$ for
simplicity) satisfies for all $s\geq -\log T'$ (and $
s<-\log\left( T'-T\right)$ if $T'
>T$) and all $z\in\mathbb{R}^N$
\begin{equation}\label{e1.7}
g(s)\theta_{s}+\beta g(s) \theta-\int_{s_{0}}^{s}g_{2}\left(\tau
\right) \Delta \theta\left( \tau \right) d\tau
-g\left(s_{0}\right) \Delta \theta_{00} =g(s)\left| \theta\right|
^{p-1}\theta
\end{equation}
where $g(s):=e^{( \beta +1) s}$ and $g_{2}(s):=e^{( \beta -1) s}$.
Denote by  $h(s):=e^{-(\beta+1)s} $ and
$h_{2}(s):=e^{-(\beta-1)s}$.

In the new set of variables $(s,z)$, the behavior of $u_t$ as
$t\uparrow T$ is equivalent to the behavior of $\theta$ as
$s\rightarrow \infty $.

In Section 2 we define an associated energy to equation
(\ref{e1.7}) that is decreasing (see Proposition \ref{l1}).
\bigskip

\noindent Our main result in this paper is:
\begin{theorem}\label{t1}\textbf{(Bounds on $\theta$)}\\
Assume that $(u_0,u_1)$ and $p$ satisfy (\ref{hypp}). If $u$ is a
blowing-up solution at time $T>0$ of (\ref{e1.1}) and $\theta$ is
defined as in (\ref{e1.5}) and satisfies
\begin{eqnarray}\label{e1.9} C\le E[\theta](s)
\end{eqnarray}
for some constant $C$ and for all $s\ge s_0$, then there exists
$K>0$ that depends on $N$, $p$ and bounds on $T$ and the initial
data in $Y_{loc,u}$ such that
$$\sup_{\begin{array}{c} s>s_{0} \\
a\in \mathbb{R}^N\end{array}} \left[ e^{-2s}\left\|h_{2}\star
\theta(s,\cdot)
 \right\| _{H^1(B)}^2+\left\| \theta(s,\cdot) \right\| _{L^2(B)}^2
 \right] <K, $$
 where $B$ is the unit ball of $\mathbb{R}^N$
and
$h_{2}\star\theta(s,z):=\int_{s_0}^sh_{2}(s-s')\theta(s',z)\,ds'$.

\end{theorem}

This can be translated in terms of $u$:
\begin{theorem}\label{p1}\textbf{(Bounds on blowing-up solutions of (\ref{e1.1}))}\\
Assume that  $(u_0,u_1)$ and $p$ satisfy (\ref{hypp}). If $u$ is a
blowing-up solution at time $T>0$ of equation (\ref{e1.1}) and
$\theta$ is defined as in (\ref{e1.5}) satisfying (\ref{e1.9}),
then there exists a positive constant $C$ , that depends on $N$,
$p$ and bounds on $T$ and the initial data in $Y_{loc,u}$, such
that for all $t\in [T(1-e^{-1}),T)$, and all $a\in \mathbb{R}^N$:
$$(T-t)^{2\beta}\left[
\| u\| _{H^2( B_{a}) }^2 + \| u_t(t) \| _{H^1(B_a) }^2\right]\leq
C,
$$
where $B_{a}$ is the unit ball centered at $a$.
\end{theorem}

In Section 3 we provide the proof of Theorem \ref{t1}. In the last
section we improve the regularity of the solution by providing a
control on the $L^r$-norm of $\theta$ and $e^{-s}[h_{2}\star
\theta]$ for $1\le r\le \frac{2N}{N-2}$.

\noindent\textbf{Acknowledgment.} This work is strongly inspired
by the series of papers of Merle and Zaag \cite{MerleZaag2003,
MerleZaag2005MA, MerleZaag2005IMRN}.

\section{The associated energy} In this section we define a
weighted energy associated to the equation (\ref{e1.7}) as
follows:
\begin{eqnarray*}
E(s)&:=&\beta/2\int_Bg(s)\rho^\alpha\theta(s)^2dz-\frac1{p+1}\int_Bg(s)\rho^\alpha|\theta(s)|^{p+1}dz\\
&&-\frac18\int_{s_{0}}^{s}\int_B\rho^\alpha
g_2(\tau)\left[|4\nabla \theta(\tau)-
\nabla\theta(s)|^2 -|\nabla\theta(s)|^2\right]dzd\tau \\
&&-
\alpha\int_{s_0}^s\int_Bg_2(\tau)[N\rho-2(\alpha-1)z^2]\rho^{\alpha-2}
\left[|\theta(s)-\theta(\tau)|^2-|\theta(s)|^2\right]dzd\tau\\
&&-\alpha\int_{s_0}^s\int_Bg(\tau)\rho^{\alpha-1}
\left[|e^{-2\tau}z\nabla
\theta(s)-\theta(\tau)|^2-|e^{-2\tau}z\nabla
\theta(s)|^2\right]dzd\tau\\
&&+\frac{1}{2}g(s_{0})\int_B\rho^{\alpha}\left[|\nabla\theta(s)+\nabla
\theta_{00}|^{2}-|\nabla\theta(s)|^{2}\right]dz\\
&&+\alpha g(s_{0})\int_B\rho^{\alpha-1}\left[|\theta(s)-z\nabla
\theta_{00}|^{2}-|\theta(s)|^{2}\right]dz,
\end{eqnarray*}
where $B$ denotes the unit ball, $\alpha$ is any number satisfying
$\alpha >\max \{ \beta ( \beta +1)/2,1+2\beta ,2\}$, and
$\rho(z):=1-|z|^2$.

In this section we prove the following
\begin{proposition}\label{l1} The energy $s\mapsto E(s)$ is a decreasing function for
$s\ge s_0$. Moreover, we have
\begin{eqnarray}
E(s+1)-E(s)&=&-\frac{\beta+1}{p+1}\int_s^{s+1}\int_B
g(s)\rho^\alpha|\theta(s')|^{p+1}dzds'\\
&&-\int_s^{s+1}\int_Bg(s)\rho^\alpha
\theta_s^2(s')dzds'\nonumber\\
& &-[\alpha-\beta(\beta+1)/2]
\int_s^{s+1}g(s')\int_B\rho^{\alpha}\theta^2(s')dzds'\nonumber\\
&&-\alpha\int_s^{s+1}\int_B
g(s')\rho^{\alpha-1}|z|^{2}|\theta(s')|^{2}dzds'\nonumber\\
&&-\int_s^{s+1}\int_B
g_{2}(s')\rho^{\alpha}|\nabla\theta(s')|^{2}dzds'.\nonumber
\end{eqnarray}
\end{proposition}
\noindent\textbf{Proof}: In order to calculate the derivative of
$E$, multiply the equation (\ref{e1.7}) by $\rho^\alpha \theta_s$
and integrate over $B:=\{|z|\le1\}$. Then we get
\begin{eqnarray}\label{e2.2}
\int_Bg(s)\rho^\alpha\left[\beta\theta\theta_s-\theta|\theta|^{p-1}\theta_s
\right]dz -\int_{s_0}^s\int_B\rho^\alpha
g_2(\tau)\theta_s(s)\Delta \theta(\tau)dzd\tau-B'&&\nonumber \\
&&\hskip-7cm=-\int_B g(s)\rho^\alpha\theta_s^2dz.
\end{eqnarray}
with
\begin{eqnarray*}
B'&:=& g(s_0)\int_B\rho^\alpha\Delta
\theta_{00}(s_0,z)\theta_s(s,z)dz\\
&=&-g(s_0)\int_B \rho^\alpha\nabla\theta_s\nabla \theta_{00}dz+
2\alpha g(s_0)\int_B\rho^{\alpha-1}z\theta_s\nabla \theta_{00}dz\\
&=&-(B_1+B_2).
\end{eqnarray*}
We have
\begin{eqnarray*}
B_{1}&=&g(s_{0})\int_B \rho^{\alpha}\nabla\theta_{s}\nabla \theta_{00}dz\\
&=&\frac{1}{2}g(s_{0})\frac{d}{ds}\int_B\rho^{\alpha}\left[|\nabla\theta(s)+\nabla
\theta_{00}|^{2}-|\nabla\theta(s)|^{2}\right]dz,
\end{eqnarray*}
and
\begin{eqnarray*}
B_{2}&=&-2\alpha g(s_{0})\int_B\rho^{\alpha-1}z\theta_{s}\nabla \theta_{00}\\
&=&\alpha
g(s_{0})\frac{d}{ds}\int_B\rho^{\alpha-1}\left[|\theta(s)-z\nabla
\theta_{00}|^{2}-|\theta(s)|^{2}\right].
\end{eqnarray*}
Using Green's formula, we write the term
$$-\int_{s_0}^s\int_B\rho^\alpha g_2(\tau)\theta_s(s)\Delta
\theta(\tau)dzd\tau=I_1+I_2$$ where
$$I_1:=\int_{s_0}^s\int_Bg_2(\tau)\rho^\alpha \nabla \theta_s(s)\nabla \theta(\tau)dzd\tau$$
and
$$I_2:=-2\alpha\int_{s_0}^s\int_Bg_2(\tau)\rho^{\alpha-1}\theta_s(s)z\nabla \theta(\tau)dzd\tau.$$
For $I_1$:
\begin{eqnarray*}
I_1&=&-\frac{1}{2}\frac d{ds}\int_{s_0}^s\int_B\rho^\alpha
g_2(\tau)|2\nabla \theta(\tau)-\frac{1}{2}\nabla \theta(s)|^2dzd\tau\\
&&+ \frac{1}{8}\frac d{ds}
\left[\int_{s_0}^sg_2(\tau)d\tau\int_B\rho^\alpha
|\nabla \theta(s)|^2dz\right]\\
&&+ g_{2}(s)\int_B \rho^{\alpha}|\nabla \theta(s)|^2dz.
\end{eqnarray*}
Remainder $I_2$:
\begin{eqnarray*}
I_2&=&-2\alpha\int_{s_0}^s\int_Bg_2(\tau)\rho^{\alpha-1}\theta_s(s)z\nabla \theta(\tau)dzd\tau\\
&=&2\alpha\int_{s_0}^s\int_Bg_2(\tau)\nabla [z\rho^{\alpha-1}\theta_s(s)]\theta(\tau)dzd\tau\\
&=&2N\alpha\int_{s_0}^s\int_Bg_2(\tau)\rho^{\alpha-1}\theta_s(s)\theta(\tau)dzd\tau\\
&&-4\alpha(\alpha-1)\int_{s_0}^s\int_Bg_2(\tau)\rho^{\alpha-2}z^2\theta_s(s)\theta(\tau)dzd\tau\\
&&+2\alpha\int_{s_0}^s\int_Bg_2(\tau)\rho^{\alpha-1}z\nabla \theta_s(s)\theta(\tau)dzd\tau\\
&=&A_1+A_2+A_3,
\end{eqnarray*}
with
\begin{eqnarray*}
A_1&:=&2N\alpha\int_{s_0}^s\int_Bg_2(\tau)\rho^{\alpha-1}\theta_s(s)\theta(\tau)dzd\tau\\
&=&-N\alpha\frac d{ds}
\left[\int_{s_0}^s\int_Bg_2(\tau)\rho^{\alpha-1}|\theta(s)-\theta(\tau)|^2dzd\tau
\right]\\
&&+N\alpha\frac d{ds}
\left[\int_{s_0}^sg_2(\tau)d\tau\int_B\rho^{\alpha-1}|\theta(s)|^2dz
\right]\\
&&-N\alpha g_2(s)\int_B\rho^{\alpha-1}|\theta(s)|^2dz,
\end{eqnarray*}
\begin{eqnarray*}
A_2&:=&-4\alpha(\alpha-1)\int_{s_0}^s\int_Bg_2(\tau)\rho^{\alpha-2}z^2\theta_s(s)\theta(\tau)dzd\tau\\
&=&2\alpha(\alpha-1)\frac d{ds}
\left[\int_{s_0}^s\int_Bg_2(\tau)z^2\rho^{\alpha-2}|\theta(s)-\theta(\tau)|^2dzd\tau
\right]\\
&&-2\alpha(\alpha-1)\frac d{ds}
\left[\int_{s_0}^sg_2(\tau)d\tau\int_Bz^2\rho^{\alpha-2}|\theta(s)|^2dz
\right]\\
&&+2\alpha(\alpha-1)
g_2(s)\int_Bz^2\rho^{\alpha-2}|\theta(s)|^2dz,
\end{eqnarray*}
\begin{eqnarray*}
A_3&:=&2\alpha\int_{s_0}^s\int_Bg_2(\tau)\rho^{\alpha-1}z\nabla \theta_s(s)\theta(\tau)dzd\tau\\
&:=&2\alpha\int_{s_0}^s\int_Bg(\tau)\rho^{\alpha-1}e^{-2\tau}z\nabla (\theta_s(s))\theta(\tau)dzd\tau\\
&=&-\alpha\frac d{ds}
\left[\int_{s_0}^s\int_Bg(\tau)\rho^{\alpha-1}|e^{-2\tau}z\nabla
\theta(s)-\theta(\tau)|^2dzd\tau
\right]\\
& &+\alpha\frac d{ds}
\left[\int_{s_0}^sg_4(\tau)d\tau\int_Bz^2\rho^{\alpha-1}|\nabla
\theta(s)|^2dz
\right]\\
& &+ \alpha\frac d{ds}
\int_{s_0}^s\int_Bg(\tau)\rho^{\alpha-1}|\theta(\tau)|^{2}dzd\tau-
2\alpha g_{2}(s) \int_B\rho^{\alpha -1}z\nabla \theta(s) \theta(s)dz\\
&=&-\alpha\frac d{ds}
\int_{s_0}^s\int_Bg(\tau)\rho^{\alpha-1}|e^{-2\tau}z\nabla
\theta(s)-\theta(\tau)|^2dzd\tau
\\
& &+\alpha\frac d{ds}
\left[\int_{s_0}^sg_4(\tau)d\tau\int_Bz^2\rho^{\alpha-1}|\nabla
\theta(s)|^2dz
\right]\\
& &+ \alpha g(s)\int_B\rho^{\alpha-1}|\theta(s)|^{2}dz- \alpha
g_{2}(s)
\int_B\rho^{\alpha -1}z\nabla \theta^{2}(s)dz\\
&=&-\alpha\frac d{ds}
\int_{s_0}^s\int_Bg(\tau)\rho^{\alpha-1}|e^{-2\tau}z\nabla
\theta(s)-\theta(\tau)|^2dzd\tau
\\
& &+\alpha\frac d{ds}
\left[\int_{s_0}^sg_4(\tau)d\tau\int_Bz^2\rho^{\alpha-1}|\nabla
\theta(s)|^2dz
\right]\\
& &+ \alpha g(s)\int_B\rho^{\alpha-1}|\theta(s)|^{2}dz+ \alpha
g_{2}(s) \int_B\nabla (\rho^{\alpha -1}z)\theta^{2}(s)dz\\
&=&-\alpha\frac d{ds}
\int_{s_0}^s\int_Bg(\tau)\rho^{\alpha-1}|e^{-2\tau}z\nabla
\theta(s)-\theta(\tau)|^2dzd\tau
\\
& &+\alpha\frac d{ds}
\left[\int_{s_0}^sg_4(\tau)d\tau\int_Bz^2\rho^{\alpha-1}|\nabla
\theta(s)|^2dz
\right]\\
& &+ \alpha g(s)\int_B\rho^{\alpha-1}|\theta(s)|^{2}dz+ \alpha N
g_{2}(s)\int_B\rho^{\alpha -1}\theta^{2}(s)dz\\
& &-2\alpha(\alpha -1)g_{2}(s)\int_B \rho^{\alpha
-2}|z|^{2}\theta^{2}(s)dz.
\end{eqnarray*}
Then
\begin{eqnarray*}
I_2&=&-\alpha\frac d{ds}
\int_{s_0}^s\int_Bg_2(\tau)[N\rho-2(\alpha-1)z^2]\rho^{\alpha-2}|\theta(s)-\theta(\tau)|^2dz
\\
&&+\alpha\frac d{ds}
\int_{s_0}^sg_2(\tau)d\tau\int_B[N\rho-2(\alpha-1)z^2]\rho^{\alpha-2}|\theta(s)|^2dz
\\
&&-\alpha\frac d{ds}
\int_{s_0}^s\int_Bg(\tau)\rho^{\alpha-1}|e^{-2\tau}z\nabla
\theta(s)-\theta(\tau)|^2dzd\tau
\\
&&+\alpha\frac d{ds}
\int_{s_0}^sg_4(\tau)d\tau\int_Bz^2\rho^{\alpha-1}|\nabla
\theta(s)|^2dz
\\
&&+\alpha g(s)\int_B\rho^{\alpha-1}|\theta(s)|^2dz.
\end{eqnarray*}
Putting $B_1$, $B_2$, $I_1$ and $I_2$ into (\ref{e2.2}) we finally
get
\begin{eqnarray}\label{energy1}
\frac d{ds}E(s)&=&-\frac{\beta+1}{p+1}\int_B
g(s)\rho^\alpha|\theta(s)|^{p+1}dz-\int_Bg(s)\rho^\alpha
\theta_s^2(s)dz\\
& &-[\alpha-\beta(\beta+1)/2]
\int_Bg(s)\rho^{\alpha}\theta^2(s)dz-\alpha\int_B
g(s)\rho^{\alpha-1}|z|^{2}|\theta(s)|^{2}\nonumber\\
& &-\int_B g_{2}(s)\rho^{\alpha}|\nabla\theta(s)|^2.\nonumber
\end{eqnarray}
which terminates the proof of the lemma. \qeda

\section{Bounds on $\theta$: Proof of Theorem \ref{t1}}

We start by the following corollary of Proposition \ref{l1}

\begin{corollary}[Bounds on $E$ and $\theta$]\label{c1}
For all $s\ge s_0$ we have
\begin{equation}\label{1}C\le E[\theta(s)]\le
E[\theta(s_0)]=: C_0,\end{equation}
\begin{equation}\label{2} \int_s^{s+1}\int_B g(s')\rho^\alpha
(\theta_s^2+|\theta|^{p+1}+\theta^2)\,dyds'+\int_s^{s+1}\int_B
g_{2}(s')\rho^\alpha|\nabla \theta|^2\le C,
\end{equation}
\begin{equation}\label{3} \int_s^{s+1}\int_B \rho^\alpha
(\theta_s^2+|\theta|^2+|\theta|^{p+1}+|\nabla \theta|^2)\,dyds'\le
C,
\end{equation}
\begin{equation}\label{4}
\int_s^{s+1}\int_{B_{1/2}}
(\theta_s^2+|\theta|^2+|\theta|^{p+1}+|\nabla \theta|^2)\,dyds'\le
C,
\end{equation}
where $C$ depends only on bounds on $T$, and the initial data of
(\ref{e1.1}) in $Y_{loc,u}$.
\end{corollary}
\noindent\textbf{Proof:} Inequalities (\ref{1}) and (\ref{2})
follow directly from Proposition \ref{l1}. Inequality (\ref{3})
follows from (\ref{2}) writing
\begin{eqnarray*}\int_s^{s+1}\int_B \rho^\alpha
(\theta_s^2+|\theta|^{p+1}+|\nabla \theta|^2+\theta^2)\,dyds'&\le& \min(h(s_{0}),h_{2}(s_0))\times \\
&&\hskip-6cm\times\left[\int_s^{s+1}\int_B g(s')\rho^\alpha
(\theta_s^2+|\theta|^{p+1}+\theta^2)\,dyds' +\int_s^{s+1}\int_B
g_{2}(s')\rho^\alpha|\nabla \theta|^2\right]\le C.\end{eqnarray*}

\noindent Similarly, since $\rho^\alpha\ge 3/4$ over $B_{1/2}$,
inequality (\ref{4}) follows from (\ref{3}). \qeda

The proof of Theorem \ref{t1} will be done in the following three
propositions:

\begin{proposition}[Control of $\theta$ in
$L^2_{loc,u}$]\label{p31} For all $s\ge s_0+1$ and all
$a\in\mathbb{R}^N$ we have
\begin{equation}\label{ep33}
\int_B \theta_a^2\,dz\le C.
\end{equation}
\end{proposition}

\begin{proposition}[Control of ${e^{-s}[h_{2}\star \nabla \theta]}$ in
$L^2_{loc,u}$]\label{p32} For all $s\ge s_0+1$ and all
$a\in\mathbb{R}^N$ we have
\begin{equation}\label{p311}e^{-2s}\int_B{\left| h_{2}\star \nabla
\theta_a(s,z)\right|}^2\,dz\le C.\end{equation}
\end{proposition}

\begin{proposition}[Control of ${e^{-s}[h_{2}\star \theta]}$ in
$L^2_{loc,u}$]\label{p33} For all $s\ge s_0+1$ and all
$a\in\mathbb{R}^N$ we have
$$
e^{-2s}\int_B [h_{2}\star \theta_a]^2\,dz\le C.
$$
\end{proposition}

\noindent\textit{Strategy of the proof:} Following
\cite{MerleZaag2005MA} and by a covering technique, we start
showing that we can insert $\rho^\alpha$ inside the integral
$\int_B$, then, using mean value theorem, we bound $\int_B$ by
$\int_s^{s+1}\int_B$. We terminate by straightforward (but tricky)
calculations using inequalities of Corollary \ref{c1}.

\noindent\textbf{Proof of proposition \ref{p31}: 1.} Let
$a_0:=a_0(s)$ be such that
$$
\int_B\rho ^\alpha \theta_{a_0}^2(s,z)\,dz\geq
\frac12\sup_{a\in\mathbb{R}^N} \int_B\rho ^\alpha
\theta_a^2(s,z)\,dz.
$$
We have:
\begin{lemma}
For all $s\ge s_0+1$ and for any $a\in\mathbb{R}^N$, we have
\begin{equation}\label{78}
\int_B \theta_{a}^2(s,z)\,dz\le C\int_B \rho^\alpha
\theta_{a_0}^2(s,z)\,dz.
\end{equation}
\end{lemma}
\noindent\textbf{Proof of Lemma 1:} Using the definition
(\ref{e1.5}) of $\theta$ and the fact that $\rho\ge 3/4$ over
$B_{1/2}$ we have
\begin{eqnarray*}
\int_{B_{\frac12}}\theta_{a}^2(z_0+z,s)\, dz &=&\int_{B_{\frac12}}
\theta_{a+z_0}^2(z,s)\, dz \\
&\le& C\int_{B_{\frac12}} \rho^\alpha \theta_{a+z_0}^2(z,s)\,dz \\
&\le& C \sup_{a\in\mathbb{R}^N} \int_B\rho ^\alpha \theta_a^2\,dz
\le 2C\int_B\rho^\alpha \theta_{a_0}^2\,dz,
\end{eqnarray*}
uniformly with respect to $z_0\in B$. Now since we can cover the
ball $B$ with $k(N)$ balls of radius $1/2$, this proves
(\ref{78}).\qeda

 \noindent\textbf{2.} Remains to prove that
$$
\int_B\rho^\alpha \theta_{a_0}^2(s,z)\,dz\le C.
$$
Using the mean value theorem and (\ref{3}), there exists $\tau\in
[s,s+1]$ such that $$\int_B\rho^\alpha
\theta_{a_0}^2(\tau,z)\,dz=\int_s^{s+1}\int_B\rho^\alpha
\theta_{a_0}^2(s',z)\,dzds'\le C.
$$
Now
\begin{eqnarray*}
\int_B\rho^\alpha \theta_{a_0}^2(s,z)\,dz&=&\int_B\rho^\alpha
\theta_{a_0}^2(\tau,z)\,dz- \int_s^\tau \int_B\rho^\alpha
\frac\partial{\partial s}[\theta_{a_0}^2](s',z)\,dzds'\\
&\le&C-\int_s^\tau \int_B\rho^\alpha
2\theta_{a_0}(\theta_{a_0})_s(s',z)\,dzds'\\
&\le&C+\int_s^\tau\int_B\rho^\alpha[\theta_{a_0}^2+(\theta_{a_0})_s^2]\,dzds'\\
&\le&C+\int_s^{s+1}\int_B\rho^\alpha[\theta_{a_0}^2+(\theta_{a_0})_s^2]\,dzds'\\
&\le&3C\qquad (\mbox{ by }(\ref{3})).
\end{eqnarray*}
This ends the proof of Proposition \ref{p31}.\qeda

\noindent\textbf{Proof of Proposition \ref{p32}: 1.} For $s\ge
s_0+1$ let $a_1=a_1(s)$ be such that
$$
e^{-2s}\int_B\rho ^\alpha \left[\int_{s_0}^s h_{2}(s-s') \nabla
\theta_{a_1}ds' \right]^2\,dz\geq \frac12\sup_{a\in\mathbb{R}^N}
e^{-2s}\int_B\rho ^\alpha \left[\int_{s_0}^s h_{2}(s-s') \nabla
\theta_ads' \right]^2\,dz.
$$
We need the following:
\begin{lemma}
For all $s\ge s_0+1$ and for any $a\in\mathbb{R}^N$, we have
\begin{equation}\label{7}
e^{-2s}\int_B [h_{2}\star \nabla \theta_{a}(s,z)]^2\,dz\le
Ce^{-2s}\int_B \rho^\alpha[h_{2}\star \nabla \theta_{a_1}]^2\,dz.
\end{equation}
\end{lemma}
\noindent\textbf{Proof of Lemma 2:} Using the definition
(\ref{e1.5}) of $\theta$ and the fact that $\rho\ge 3/4$ over
$B_{1/2}$ we have
\begin{eqnarray*}
e^{-2s}\int_{B_{\frac12}}\left[ \int_{s_0}^sh_{2}(s-s') \nabla
\theta_{a}(z_0+z,s') ds' \right]^2dz &&\\
&&\hskip-4cm=e^{-2s}\int_{B_{\frac12}} \left[
\int_{s_0}^sh_{2}(s-s')\nabla
\theta_{a+z_0}(z,s') ds' \right]^2dz \\
&&\hskip-4cm\le Ce^{-2s}\int_{B_{\frac12}} \rho^\alpha\left[
\int_{s_0}^sh_{2}(s-s')\nabla \theta_{a+z_0}(z,s') ds' \right]^2dz \\
&&\hskip-4cm\le C \sup_{a\in\mathbb{R}^N} e^{-2s}\int_B\rho
^\alpha
\left[\int_{s_0}^s h_{2}(s-s') \nabla \theta_a(z,s')ds' \right]^2\,dz\\
&&\hskip-4cm\le Ce^{-2s}\int_B\rho^\alpha\left[
\int_{s_0}^sh_{2}(s-s') \nabla \theta_{a_1}(z,s') ds' \right]^2dz,
\end{eqnarray*}
uniformly with respect to $z_0\in B$. Now since we can cover the
ball $B$ with $k(N)$ balls or radius $1/2$, this proves
(\ref{7}).\qeda

\noindent\textbf{2.} Now we will prove that
\begin{equation}\label{i2}
\int_s^{s+1}e^{-2s'}\int_B\rho^\alpha [h_{2}\star\nabla
\theta_{a_1}]^2(s',z)\,ds'dz\le C.
\end{equation}
By integration by parts we have
\begin{eqnarray}\label{e30}
\int_B\rho^\alpha\Delta
\theta(s',z)\theta(s,z)\,dz&=&-\int_B\rho^\alpha \nabla
\theta(s',z)\nabla
\theta(s,z)\,dz\\
&&\hskip1cm+2\alpha\int_B\rho^{\alpha-1}\theta(s,z)z\cdot\nabla
\theta(s',z)\,dz.\nonumber
\end{eqnarray}
Thus
\begin{eqnarray}\label{e31}
\int_s^{s+1}e^{-2s'}\int_B \rho^\alpha \theta[h_{2}\star \Delta
\theta]\,dzds'&=&-\int_s^{s+1}e^{-2s'}\int_B\rho^\alpha[h_{2}\star
\nabla
\theta]\cdot\nabla \theta\,dzds'\\
&&\hskip.5cm+2\alpha\int_s^{s+1}e^{-2s'}\int_B\rho^{\alpha-1}z\cdot[h_{2}\star\nabla
\theta]\theta\,dyds'.\nonumber
\end{eqnarray}
Now, since $$\frac{\partial}{\partial s}[e^{-s}(h_{2}\star
f)]=e^{-s}[f-\beta(h_{2}\star f)],$$ so, for $s_1<s_2$ we have
\begin{equation}\label{e31bis}\frac12\left[e^{-2s'}|h_{2}\star \nabla
\theta|^2\right]_{s_1}^{s_2} =\int_{s_1}^{s_2}e^{-2s'}[h_{2}\star
\nabla \theta]\cdot \nabla \theta\,ds'
-\beta\int_{s_1}^{s_2}e^{-2s'}|h_{2}\star \nabla
\theta|^2\,ds'.\end{equation} Multiplying equation (\ref{e1.7}) by
$\rho^\alpha \theta_{a_1}$ and then integrating over
$[s,s+1]\times B$ we get (using (\ref{e30}), (\ref{e31}) and
(\ref{e31bis}))
\begin{eqnarray*}
\frac12\int_B\rho^\alpha\left[e^{-2s'}|h_{2}\star \nabla
\theta_{a_1}|^2 \right]_{s}^{s+1}\,dz +\beta\int_{s}^{s+1}\int_B
\rho^\alpha e^{-2s'}|h_{2}\star \nabla \theta_{a_1}|^2\,dzds'&&\\
-2\alpha\int_s^{s+1}\int_B\rho^{\alpha-1}e^{-2s'}z\cdot[h_{2}\star\nabla
\theta_{a_1}]\theta_{a_1}\,dzds'.&&\\
&&\hskip-11cm=-\int_s^{s+1}\int_B\rho^\alpha
\theta_{a_1}\left[(\theta_{a_1})_s+\beta
\theta_{a_1}-h(s-s_0)\Delta
\theta_{00}-|\theta_{a_1}|^{p+1}\right]dzds'.\\
\end{eqnarray*}
Using the inequality $\pm ab\le\gamma^{-1}a^2+\frac\gamma4b^2$, we
have
$$\int_s^{s+1}\int_B\rho^{\alpha-1}ze^{-2s'}[h_{2}\star\nabla \theta]\,\theta\le
\gamma^{-1}\int_s^{s+1}\int_B\rho^\alpha e^{-2s'}|h_{2}\star\nabla
\theta|^2+\frac\gamma4\int_s^{s+1}\int_B\rho^{\alpha-2}|z|^2\theta^2,
$$
where $\gamma =4\frac\alpha\beta$. Then, using (\ref{3}) and
proposition \ref{p31}, we get
\begin{equation}\label{e32}
\frac\beta 2\int_s^{s+1}\int_B \rho^\alpha e^{-2s'}|h_{2}\star
\nabla \theta_{a_1}|^2\,dzds'
+\frac12\int_B\left[e^{-2s'}|h_{2}\star \nabla w_{a_1}|^2
\right]_{s}^{s+1}\,dy\le C.
\end{equation}
This can be written as
$$y'(s)+ay(s)\le b,$$
where $a$ and $b$ are positive constants and
$$y(s):=\int_s^{s+1}\int_B \rho^\alpha e^{-2s'}|h_{2}\star \nabla
\theta_{a_1}|^2\,dzds'.$$ This directly gives (\ref{i2}).

\noindent\textbf{3.} Remains to prove that, for all $s\ge s_0+1$,
we have
\begin{equation}\label{e28}
\int_B\rho^\alpha e^{-2s}|h_{2}\star\nabla
\theta_{a_1}|^2(s,z)\,dz\le C.
\end{equation}
Using the mean value theorem and (\ref{3}), there exists $\tau\in
[s,s+1]$ such that
$$\int_B\rho^\alpha e^{-2\tau}[h_{2}\star\nabla \theta_{a_1}]^2(\tau,z)\,dz=
\int_s^{s+1}\int_B\rho^\alpha e^{-2s'}[h_{2}\star\nabla
\theta_{a_1}]^2(s',z)\,dzds'\le C.$$ Then
\begin{eqnarray*}
\int_B\rho^\alpha e^{-2s}[h_{2}\star\nabla
\theta_{a_1}]^2(s,z)\,dz&=&\int_B\rho^\alpha e^{-2\tau}
[h_{2}\star\nabla \theta_{a_1}]^2(\tau,z)\,dz\\ && + \int_s^\tau
\int_B\rho^\alpha \frac\partial{\partial s}(e^{-2s'}
[h_{2}\star\nabla \theta_{a_1}]^2(s',z))\,ds'dz\\
&\le&C+2\int_s^\tau \int_B\rho^\alpha e^{-2s'}
[h_{2}\star \nabla \theta_{a_1}][\nabla \theta_{a_1}-\beta h_{2}\star \nabla \theta_{a_1}]\,ds'dz\\
&\le&C+2\int_s^\tau\int_B\rho^\alpha e^{-2s'}(h_{2}\star \nabla
\theta_{a_1})\nabla \theta_{a_1}\\
&&-2\int_s^\tau\int_B\rho^\alpha \beta e^{-2s'}|h_{2}\star \nabla \theta_{a_1}|^2\}\,ds'dz\\
&\le&C+C_{1}\int_s^{s+1}\int_B\rho^\alpha e^{-2s'}|h_{2}\star
\nabla \theta_{a_1}|^2\\
&&+C_{2}\int_s^{s+1}\int_B\rho^\alpha\nabla \theta_{a_1}^2\,ds'dz\\
&\le&C'\qquad (\mbox{ by }(\ref{3})).
\end{eqnarray*}
This ends the proof of Proposition \ref{p32}.\qeda

\noindent\textbf{Proof of proposition \ref{p33}:} The proof is
similar to the previous one.\qeda

\section{Improvement of the regularity to $L^r$, $1\le r\le
\frac{2N}{N-2}$}

We terminate with an  improvement of the control on $\theta$ and
$e^{-s}[h_{2}\star \theta]$ we obtained in Propositions \ref{p31}
and \ref{p33}. In fact, using Sobolev's embedding Theorem and the
covering technique used in Propositions \ref{p31}, \ref{p32} and
\ref{p33} we can show the following:

\begin{proposition}[Control of $\theta$ and ${e^{-s}[h_{2}\star \theta]}$ in
$L^r(B)$ for $1\le r\le\frac{2N}{N-2}$]\label{p34}

Let $1\le r\le \frac{2N}{N-2}$. For all $s\ge s_0+1$ and all
$a\in\mathbb{R}^N$ we have
\begin{equation}\label{p342} e^{-rs}\int_B{\left|(h_{2}\star
\theta_a(s,z))\right|}^r\,dz\le C.\end{equation} If, in addition,
$r\le \frac {2N}{N-1}$ then
\begin{equation}\label{p341}\int_B{\left|
\theta_a(s,z)\right|}^r\,dz\le C.\end{equation}
\end{proposition}

\noindent\textbf{Proof of Proposition \ref{p34}:} The inequality
(\ref{p342}) is direct using propositions \ref{p32}, \ref{p33} and
Sobolev's injection Theorem: $H^1(B) \hookrightarrow L^r(B)$.

\noindent For the inequality (\ref{p341}) and following the proof
of Proposition \ref{p31}, let $a_3:=a_3(s)$ be such that
$$
\int_B\rho ^\alpha \theta_{a_3}^r(s,z)\,dz\geq
\frac12\sup_{a\in\mathbb{R}^N} \int_B\rho ^\alpha
\theta_a^r(s,z)\,dz,
$$
where $\theta_{a_3}^r$ stand for $|\theta_{a_3}|^r$. Similarly, we
get:
\begin{equation}\label{781}
\int_B \theta_{a_3}^r(s,z)\,dz\le C\int_B \rho^\alpha
\theta_{a_3}^r(s,z)\,dz.
\end{equation}
Using the mean value theorem and (\ref{3}), there exists $\tau\in
[s,s+1]$ such that
$$\int_B\rho^\alpha \theta_{a_3}^r(\tau,z)\,dz=\int_s^{s+1}\int_B\rho^\alpha \theta_{a_3}^r(s',z)\,dzds'\le C.$$
Now, using Sobolev's embedding theorem $H^1 ((s,s+1)\times B)
\hookrightarrow L^{2(r-1)}((s,s+1)\times B)$, we get
\begin{eqnarray*}
\int_B\rho^\alpha \theta_{a_3}^r(s,z)\,dz&=&\int_B\rho^\alpha
\theta_{a_3}^r(\tau,z)\,dz+\int_s^\tau \int_B\rho^\alpha
\frac\partial{\partial s}\theta_{a_3}^r(s',z)\,ds'dz\\
&\le&C+\int_s^\tau \int_B\rho^\alpha
r|\theta_{a_3}|^{r-1}|(\theta_{a_3})_s|(s',z)\,ds'dz\\
&\le&C+\frac12\int_s^\tau\int_B\rho^\alpha[|\theta_{a_3}|^{2(r-1)}+(\theta_{a_3})_s^2]\,ds'dz\\
&\le&C+C\left[\int_s^{s+1}\int_B\rho^\alpha[\theta_{a_3}^2+|\nabla
\theta_{a_3}|^2]\,
ds'dz\right]^{r-1}\\
&&\hskip1cm+\frac12\int_s^{s+1}\int_B\rho^\alpha(\theta_{a_3})_s^2\,ds'dz\\
&\le&C\qquad (\mbox{ by }(\ref{3})).
\end{eqnarray*}
This ends the proof of Proposition \ref{p34}.\qeda

\bibliographystyle{plain}
\bibliography{c:/bib/mybib}

\end{document}